\documentclass{evn2024}

\usepackage{natbib}

\usepackage{tikz} 
\usetikzlibrary{shapes,arrows} 
\usetikzlibrary{arrows.meta}
\tikzset{every node/.style={inner sep=2pt}}

\definecolor{lightblue}{rgb}{0.145,0.6666,1} 

\usepackage{url}
\usepackage{subfigure}

\begin{document}
   \title{
      Mitigating Source Structure in Geodetic VLBI\\
      on the Visibility Level
   }

   \author{
      F.~Jaron\inst{1,2} \and
      L.~Baldreich\inst{1}\and
      J.~B\"ohm\inst{1}\and
      P.~Charlot\inst{3}\and
      A.~Collioud\inst{3}\and
      J.~Gruber\inst{4}\and
      H.~Kr\'asn\'a\inst{1}\and
      I.~Mart\'i-Vidal\inst{5,6}\and
      A.~Nothnagel\inst{1}\and
      V.~Pérez-D\'iez\inst{7,8}
   }

   \institute{
        Technische Universit\"at Wien, Wiedner Hauptstra{\ss}e 8-10, 1040 Wien, Austria, \texttt{Frederic.Jaron@tuwien.ac.at}
        \and
        Max-Planck-Institut f\"ur Radioastronomie, Auf dem H\"ugel 69, 53121 Bonn, Germany
        \and
      Laboratoire d'Astrophysique de Bordeaux, Univ. Bordeaux, CNRS, B18N, all\'ee Geoffroy Saint-Hilaire, 33615, Pessac, France
      \and
      Federal Office of Metrology and Surveying, Austria
      \and
      Dpt. Astronomia i Astrof\'isica, Universitat de Val\`encia, C/ Dr. Moliner 50, 46120 Burjassot, Spain
      \and
      Observatori Astron\`omic, Universitat de Val\`encia, C/ Cat. José Beltrán 2, 46980 Paterna, Spain
      \and
      Observatorio Astron\'omico Nacional (OAN-IGN), Alfonso XII 3, 28014 Madrid, Spain
      \and
      Centro de Desarrollos Tecnol\'ogicos, Observatorio de Yebes (IGN), 19141 Yebes, Guadalajara, Spain
   }

   \abstract{
      Geodetic and astrometric VLBI has entered a new era with the implementation of the VLBI Global Observing System (VGOS). These broadband and dual linear polarization observations aim at an accuracy of station coordinates of 1~mm and a reference frame stability of 0.1~mm/year. Although the extended brightness distribution of many of the radio-loud active galactic nuclei observed during geodetic VLBI sessions is resolved by the interferometer, the established processing chain still treats these objects as point sources. We investigate the impact of source structure on the visibility level and develop tools to remove the structure from the visibility data, right after correlation. Here we present our approach and show results obtained from observational VGOS data.
   }

   \maketitle
%

\section{Introduction}

The geodetic and astrometric application of VLBI \citep{Sovers1998} is powerful to establish the International Celestial Reference Frame \citep[ICRF,][]{Charlot2020}, and it also contributes significantly to the International Terrestrial Reference Frame \citep[ITRF,][]{Altamimi2023} by providing the scale parameter. The radio sources observed during VLBI sessions used for this kind of analysis are radio-loud active galactic nuclei (AGN), in which the radio emission originates from a jet, which is often resolved by the interferometer. Nevertheless, in the traditional approach, all of the observations are treated as if these objects were point sources. There are studies which indicate that source structure is one of the dominating error sources in geodetic VLBI \citep{Anderson2018}.

With the implementation of the VGOS system \citep{Petrachenko2009, Niell2018}, geodetic VLBI has entered a new era. Observations with the VGOS network are carried out with a broad bandwidth and with dual linear polarization, aiming at a station position accuarcy of 1~mm and a reference frame stability of 0.1~mm/year. Systematic effects, such as source structure, are more significant for VGOS observations than they have been for legacy geodetic VLBI at S- and X-band, and would need to be corrected for in order to meet these accuracy goals. 

We have developed an approach to remove the impact of source structure from the output of the VLBI correlator, i.e., from the array of interferometer phases. Here we present our methods and show results obtained from real VGOS observations. In our project we have a focus on VGOS, but our scheme is not limited to this case.

\section{Methods}

Following the formalism developed in \citet{Charlot1990}, the structure phase is given by
\begin{equation}\label{eq:phase}
   \phi_{\rm s} = \frac{2\pi}{\lambda}\mathbf{B}\cdot\mathbf{OP_0} + \tan^{-1}\left(\frac{-Z_{\rm s}}{Z_{\rm c}}\right),
\end{equation}
where $Z_{\rm s}$ and $Z_{\rm c}$ are the two-dimensional sine- and cosine-transforms of the source brightness distribution, $\mathbf{B}$ is the baseline vector, $\mathbf{O}$ the origin of a local coordinate system in the plane of the sky near the source, $\mathbf{P_0}$ the reference point within the source, and $\lambda$ the observing wavelength. Analytical expressions for $Z_{\rm c}$ and $Z_{\rm s}$ for the case of modeling the brightness distribution with Gaussian functions are given in Eqs~(17) and~(18) of \citet{Charlot1990}.

\begin{figure}
  \begin{tikzpicture}
    \node[draw, fill=red!25!white] (obs) at (3.1,5) {Observations};
    \node[draw, fill=red!25!white] (sim) at (3.9,4) {Simulation (VieRDS)};
    \node[draw, fill=blue!25!white] (raw) at (0,5) {Raw data (level-0)};
    \node[draw, fill=red!25!white] (corr) at (0,4) {Correlation (DiFX)};
    \node[draw, fill=blue!25!white] (level1) at (0,3) {Visibilities (level-1)};
    \node[draw, fill=yellow] (viesoft) at (0,2) {\large VieSOFT};
    \node[draw, fill=blue!25!white] (level1corr) at (0,1) {Corrected visibilities (level-1)};
    \node[draw, fill=red!25!white] (ff) at (0,0) {Fringe-Fitting (HOPS fourfit)};
    \node[draw, fill=blue!25!white] (vgosdb) at (0,-1) {VGOSDB};
    \node[draw, fill=red!25!white] (analysis) at (0,-2) {Geodetic Analysis (VieVS)};
    \node[draw, fill=blue!25!white] (results) at (4.1,-2) {Results (Figs 4-6)};
    \node[draw, fill=blue!25!white] (mdl) at (3.5,2) {Source structure models};
    \node[draw, fill=red!25!white] (cal) at (4,1) {Calibration};
    \node[draw, fill=blue!25!white] (image) at (4,0) {Image (Fig.\,\ref{fig:image})};
    \draw[-latex] (raw) -- (corr);
    \draw[-latex] (corr) -- (level1);
    \draw[-latex] (level1) -- (viesoft);
    \draw[-latex] (mdl) -- (viesoft);
    \draw[-latex] (viesoft) -- (level1corr);
    \draw[-latex] (level1corr) -- (ff);
    \draw[-latex] (ff) -- (vgosdb);
    \draw[-latex] (vgosdb) -- (analysis);
    \draw[-latex] (level1corr) -- (cal);
    \draw[-latex] (cal) -- (image);
    \draw[-latex] (sim.north west) -- (raw.south east);
    \draw[-latex] (obs) -- (raw);
    \draw[-latex] (analysis) -- (results);
  \end{tikzpicture}
  \caption{
    Flowchart of our data processing pipeline, from raw data to images and geodetic parameters.
  }
  \label{fig:methods}
\end{figure}
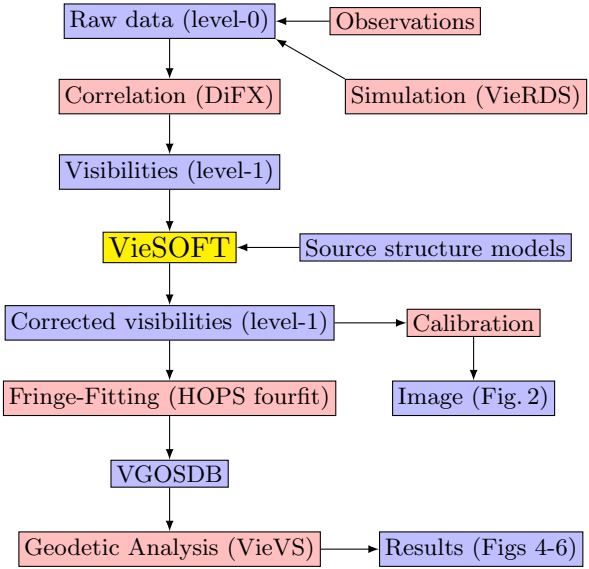

A flowchart of the processing pipeline that we use for our investigation is shown in Fig.~\ref{fig:methods}. Raw data can be obtained from real observations and simulations. Simulated baseband VLBI data can be generated using the software VieRDS \citep{Gruber2021}, which has been augmented to also simulate observations to extended sources. For the results presented here we used real VGOS observations exclusively, using the frequency setup described in \citet{Niell2018}. The data were correlated using the DiFX software package \citep{Deller2011} in its version 2.5.4, which is the current version to be used for VGOS data (as described in \citealt{Jaron2021}). DiFX outputs the visibility data in the so-called SWIN file format.

We have developed a software package \texttt{VieSOFT}, which removes structure phases from the SWIN files. It takes as configuration a description of the source brightness distribution in the form of Gaussian components. Each component refers to its own frequency range. The tool then computes $\phi_{\rm s}$ (Eq.~1 here), evaluating Eqs~(17) and~(18) in \citet{Charlot1990}, for every spectral channel in every observation for every polarization product contained in the SWIN file, and subtracts $\phi_{\rm s}$ from the original structure phases. The amplitudes of the visibilities are left unchanged. The corrected visibilities are written to SWIN files, which can be processed in the common way.

In order to check for any remaining structure in the corrected visibilities, we make use of the VGOS calibration and imaging procedure that has recently been published by \citet{Perez-Diez2024}. We fringe-fit the data with the Haystack Observatory Postprocessing System (HOPS)\footnote{\url{https://www.haystack.mit.edu/haystack-observatory-postprocessing-system-hops/}}, export them in VGOSDB format\footnote{Using nuSolve, available from \url{https://sourceforge.net/projects/nusolve/}} and perform a full geodetic analysis of the group delays using the Vienna VLBI and Satellite Software (VieVS, \citealt{Boehm2018}). The ultimate benchmark for any source structure correction should be the accuracy of the estimated geodetic parameters and the distribution of post-fit residuals.

\section{Results}

Here we describe the different ways in which we tested our approach and present our results.

\subsection{Imaging} \label{sec:image}

\begin{figure}
    \subfigure[Original]{
        \includegraphics[width=0.47\linewidth]{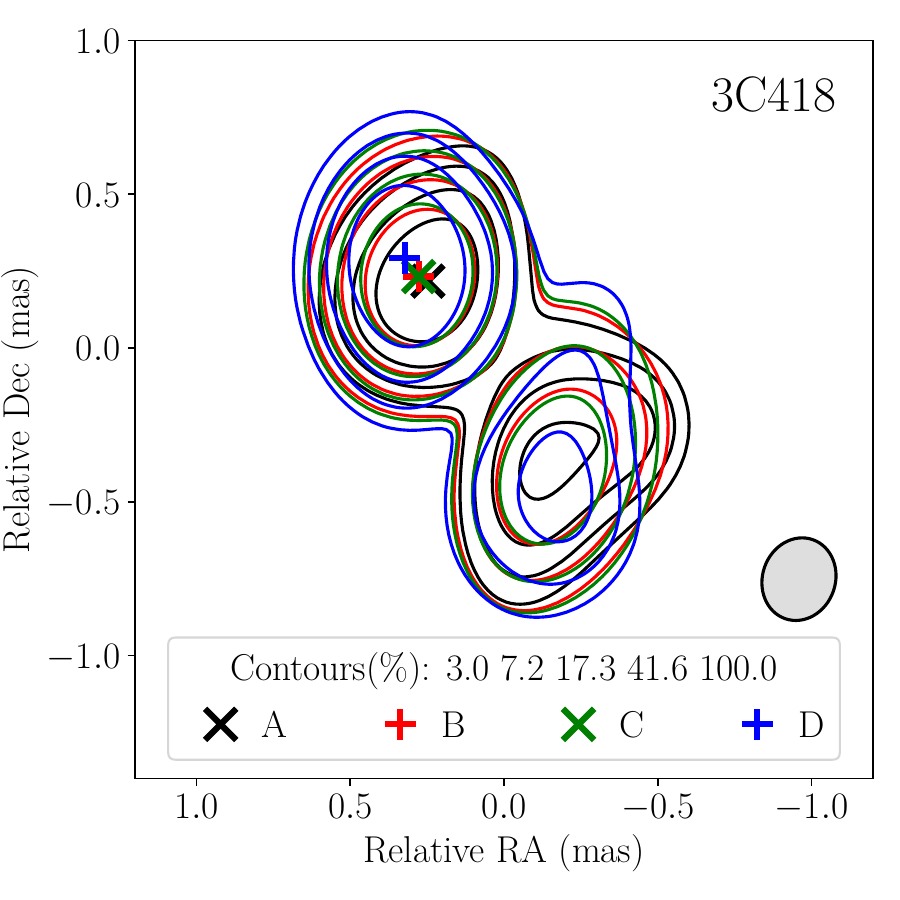}
    }
    \hfill
    \subfigure[Corrected]{
        \includegraphics[width=0.47\linewidth]{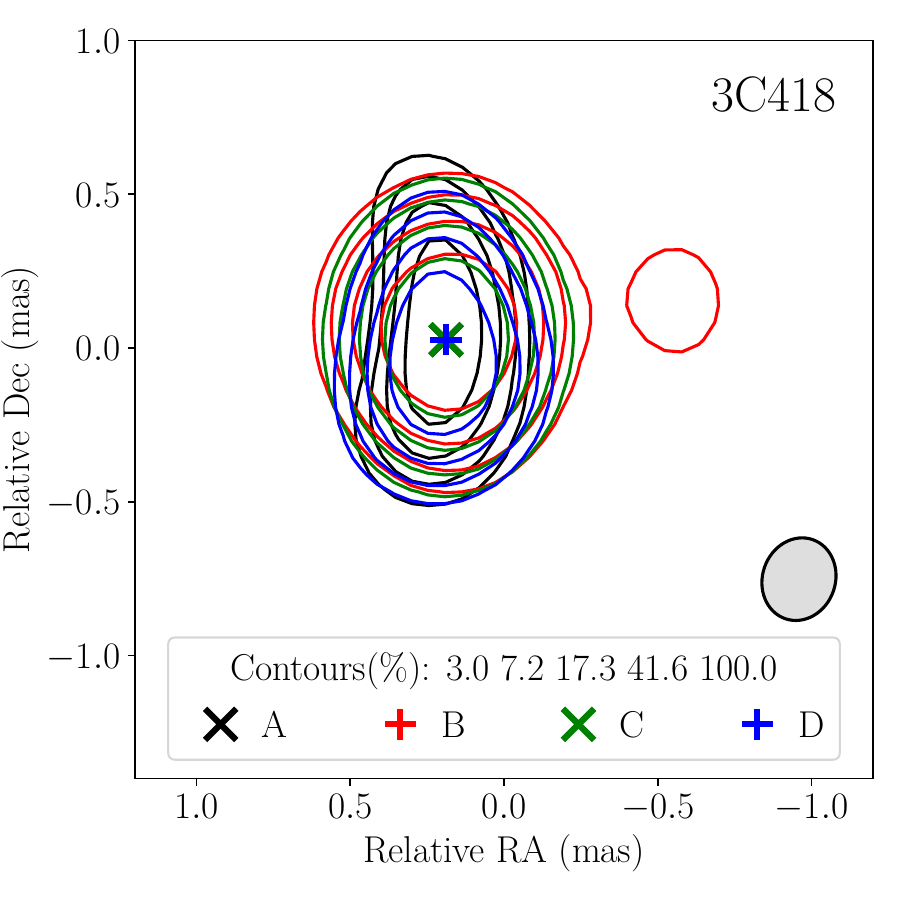}
    }
    \caption{
      (a) Original image of 3C418, obtained from analyzing VGOS session vo2187. This figure was first published in \citet{Perez-Diez2024}.
      (b) Image of the same source after removing the structure phases from the observational data of the same session.
    }
    \label{fig:image}
\end{figure}

Figure~\ref{fig:image} demonstrates the effect of our source structure correction approach in the image domain. The left panel of the figure shows the original image of the source 3C418, as obtained from analyzing VGOS observations of the session vo2187 (observed on July 6, 2022) by \citet{Perez-Diez2024}. The contours are shown in different colors, each corresponding to a certain radio frequency range: 3.0-3.5~GHz (band~A) in black, 5.2-5.7~GHz (band~B) in red, 6.3-6.8~GHz (band~C) in green, and 10.2-10.7~GHz (band~D) in blue. The position of the core component is shown by the $\times$ and + symbols in the respective colors. The source has a two-component structure, and the core position presents a dependency on frequency. We derived a model of frequency-dependent Gaussian components from this image. The image shown in the right panel is derived from the corrected visibilities. In this image the structure is greatly removed and reduced to only one component, and also the core position is the same for all frequencies. Because these are self-calibrated images, the origin of the coordinate system does not necessarily coincide with the ICRF3 position of the source.

\subsection{Effect of changing the reference point}

\begin{figure}
    \centering
   \includegraphics[width=.61\linewidth]{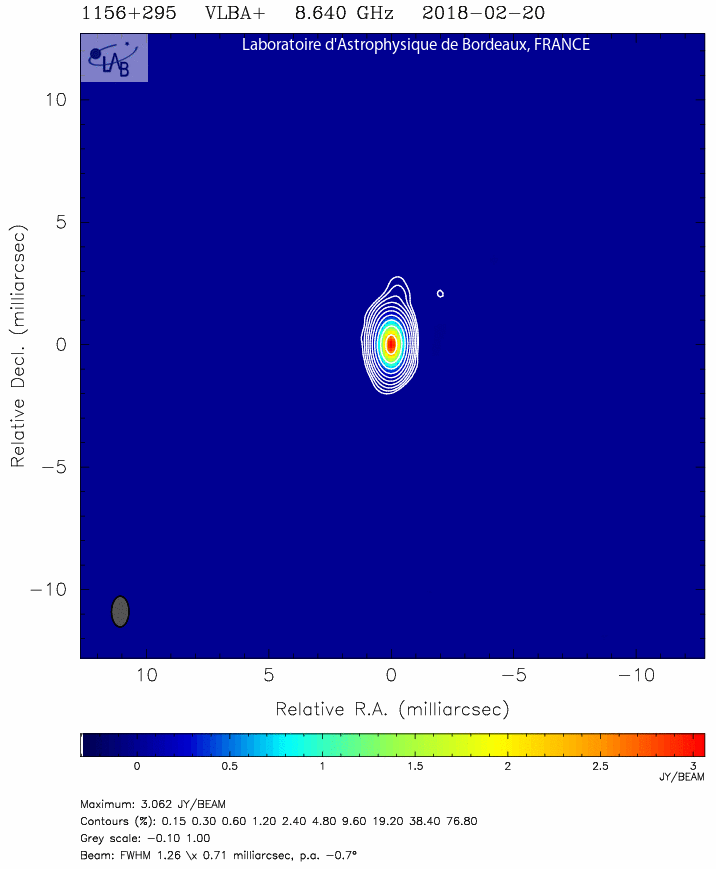}
    \caption{
      VLBI image of 1156+295 at X-band.
    }
    \label{fig:1156+295}
\end{figure}

\begin{figure}
   \includegraphics[width=.49\linewidth]{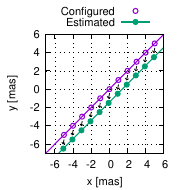}
   \hfill
   \includegraphics[width=.49\linewidth]{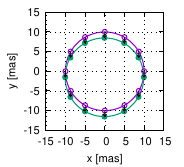}
   \caption{
        Configured reference points (purple open circles) lead to estimated source positions (green filled circles) for 1156+295, as indicated by the arrows. The estimated source positions are relative to the ICRF3 S/X position of 1156+295. In the left panel we changed the reference point along a straight line, for the right panel we did the same along a circle.
   }
   \label{fig:ref}
\end{figure}

We explored the effect that changing the reference point~$\mathbf{P_0}$ in Eq.~(\ref{eq:phase}) has on the estimated source position. For this purpose, we identified vo3012 (observed on January 12, 2023) as a well-behaved VGOS session in the geodetic sense and 1156+295 as a compact source with many observations within that session. Figure~\ref{fig:1156+295} presents an image of this source at X-band, showing that it has negligible source structure. This image is taken from the Bordeaux VLBI Image Database\footnote{ \url{https://bvid.astrophy.u-bordeaux.fr}}. We configured VieSOFT to only change the reference point for this one source in a systematic way.

We ran a full geodetic analysis on the resulting database of group delays. Figure~\ref{fig:ref} shows the estimated source positions (green) in comparison to the configured reference points (purple), relative to the ICRF3 S/X position of 1156+295. The black arrows in the figures point from configured to estimated position. In the left panel we changed the reference point in a linear fashion. Besides a systematic offset between a priori and estimated position, recognizable by comparing the configured position at the coordinate origin to the estimated one, the estimated postion follows precisely the configured ones. The same is true for the circular configuration shown in the right panel.

\subsection{Closure delays} \label{sec:closure}

\begin{figure}
   \includegraphics[width=.49\linewidth]{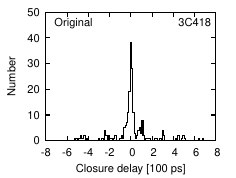}
   \hfill
   \includegraphics[width=.49\linewidth]{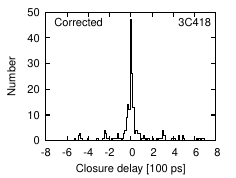}
   \caption{
      Histogram of the triangular closure delays from observations of 3C418 in the VGOS session vo2187.
      Left: Original data without any correction.
      Right: Structure phases removed.
   }
   \label{fig:closure}
\end{figure}

Consider a triangle of VLBI stations~A, B, and~C. The \emph{closure delay} is defined as the sum of the group delays,
\begin{equation}
  \tau_{\rm ABC} = \tau_{\rm AB} + \tau_{\rm BC} + \tau_{\rm CA},
\end{equation}
where all of the group delays~$\tau$ entering this sum have to refer to the same wavefront. All station-specific error sources cancel out in this equation, so in the absence of any baseline-specific systematics, the closure delays should be distributed around zero, depending on their measurement noise. Source structure is one such systematic effect that can be the reason for significant non-zero closure delays.

We investigated the closure delays resulting from observations of the source 3C418 in the VGOS session vo2187. The left panel of Fig.~\ref{fig:closure} shows the closure delay distribution resulting from the group delays obtained from the original data. The right panel shows the distribution of the closure delays after applying the same correction to the visibility data as we did in Sect.~\ref{sec:image}. Comparing the two panels in Fig.~\ref{fig:closure}, the distribution of the corrected data has a higher and narrower peak, and many of the misclosures are removed.

\subsection{Full geodetic analysis}

\begin{figure*}
   \subfigure[Original]{
    \includegraphics[width=.49\linewidth, trim={0 3.1cm 0 0}, clip]{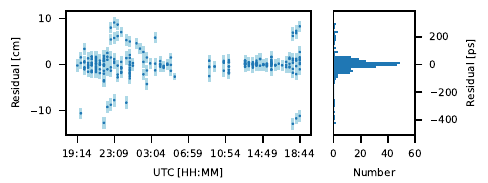}
    }
    \hfill
    \subfigure[Corrected]{
    \includegraphics[width=.49\linewidth, trim={0 3.1cm 0 0}, clip]{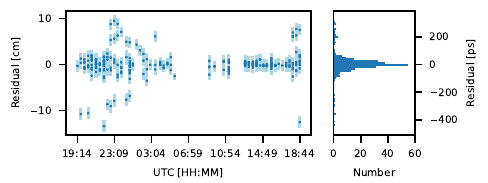}
    }
    \caption{
      Post-fit residuals after running a full geodetic solution of vo3012. Here only data resulting from observations of 3C418 are shown.
      (a) Residuals of the original data.
      (b) Residuals of the corrected data.
    }
    \label{fig:res}
\end{figure*}

We ran a full geodetic parameter estimation by analyzing group delay databases with the VLBI component of VieVS. We use a parametrization in which we estimate source coordinates, UT1-UTC, and tropospheric zenith delays. For this analysis we again make use of the session vo3012. We run the same analysis once on a database containing group delays obtained from the original observations and then again on a database of group delays obtained from the corrected visibilities.

The post-fit residuals (i.e., observed minus computed group delays) for observations of 3C418 are shown in Fig.~\ref{fig:res}. In the left-hand panel of each of the sub-figures~(a) and~(b) the residuals are expressed in centimeters (assuming a signal propagation at the speed of light) and plotted against time of the observation. The right-hand panel shows a histogram of the same residuals, expressed in picoseconds. The effect of the structure correction becomes visible when comparing the two histograms of panels~(a) and~(b). The peak for the corrected data is higher and narrower than the one for the original data. This finding is in agreement with the result that we have obtained for the analysis of the closure delays of the same session and the same source (see Fig.~\ref{fig:closure} and Sect.~\ref{sec:closure}).

\section{Conclusions}

We have developed a method for the application of source structure correction to VLBI visibility data. The data format that is currently supported is the output of the DiFX software correlator (SWIN files, \citealt{Deller2011}). The corrected data can be processed by any software for calibration and fringe-fitting, the same as used with the original SWIN files. In the ideal case the corrected data look as if they had been obtained by observations of point sources. Having tested our approach we conclude:
\begin{itemize}
  \item{
    A sample image obtained from structure-corrected visibilities indeed appears more compact than the original image (Fig.~\ref{fig:image}).
  }
  \item{
    Deliberate shifts of the reference point in the visibilities result in strictly corresponding shifts in the estimates of source positions  (Fig.~\ref{fig:ref}).
  }
  \item{
    Closure delays show a more compact distribution, but not all misclosures are removed (Fig.~\ref{fig:closure}).
  }
  \item{
    After a geodetic analysis post-fit residuals seem slightly improved (Fig.~\ref{fig:res}).
  }
\end{itemize}

The results from our tests show that it is possible to correct the source structure in visibility data, which is an early point in the VLBI processing chain, right after the correlation of the raw data. This has the benefit that already the fringe-fitting step is free from any unwanted source structure contribution. Not only the group delays, estimated in the fringe-fitting step, are thus corrected, but also other parameters such as the delay rates and, in the case of VGOS, the estimates of the differential total electron content (dTEC).

We plan to make the code applicable to other visibility formats (e.g., FITS-IDI) and make the software publicly available. Images from VLBI sessions that are both suitable for imaging \emph{and} geodesy \citep{Schartner2023} will help to improve the structure correction.

\begin{acknowledgements}
We thank Jan~Wagner of the MPIfR for reading the manuscript and providing useful feedback.
FJ, JB, and HK acknowledge funding by the Austrian Science Fund (FWF) [P 35920].
The computational results presented have been achieved in part using the Vienna Scientific Cluster (VSC).
This research made use of data obtained from observations carried out by the International VLBI Service (IVS; \citealt{Nothnagel2017}).
This work made use of the Swinburne University of Technology software correlator, developed as part of the Australian Major National Research Facilities Programme and operated under licence.
The data were correlated at the correlator of the MPIfR in Bonn, Germany. 

\end{acknowledgements}



\begin{thebibliography}{}
   \bibitem[\protect\citeauthoryear{Altamimi et al.}{2023}]{Altamimi2023} Altamimi Z., Rebischung P., Collilieux X., M{\'e}tivier L., Chanard K., 2023, JGeod, 97, 47. doi:10.1007/s00190-023-01738-w
   \bibitem[\protect\citeauthoryear{Anderson \& Xu}{2018}]{Anderson2018} Anderson J.~M., Xu M.~H., 2018, JGRB, 123, 10, 162, 190. doi:10.1029/2018JB015550
   \bibitem[\protect\citeauthoryear{B{\"o}hm et al.}{2018}]{Boehm2018} B{\"o}hm J., B{\"o}hm S., Boisits J., Girdiuk A., Gruber J., Hellerschmied A., Kr{\'a}sn{\'a} H., et al., 2018, PASP, 130, 044503. doi:10.1088/1538-3873/aaa22b
   \bibitem[\protect\citeauthoryear{Charlot}{1990}]{Charlot1990} Charlot P., 1990, AJ, 99, 1309. doi:10.1086/115419
   \bibitem[\protect\citeauthoryear{Charlot et al.}{2020}]{Charlot2020} Charlot P., Jacobs C.~S., Gordon D., Lambert S., de Witt A., B{\"o}hm J., Fey A.~L., et al., 2020, A\&A, 644, A159. doi:10.1051/0004-6361/202038368
   \bibitem[\protect\citeauthoryear{Deller et al.}{2011}]{Deller2011} Deller A.~T., Brisken W.~F., Phillips C.~J., Morgan J., Alef W., Cappallo R., Middelberg E., et al., 2011, PASP, 123, 275. doi:10.1086/658907
   \bibitem[\protect\citeauthoryear{Gruber, Nothnagel, \& B{\"o}hm}{2021}]{Gruber2021} Gruber J., Nothnagel A., B{\"o}hm J., 2021, PASP, 133, 044503. doi:10.1088/1538-3873/abeca4
   \bibitem[\protect\citeauthoryear{Jaron et al.}{2021}]{Jaron2021} Jaron F., Bernhart S., B{\"o}hm J., Gonz{\'a}lez Garc{\'\i}a J., Gruber J., Choi Y.~K., Mart{\'\i}-Vidal I., et al., 2021, Proceedings of the 25th European VLBI Group for Geodesy and Astrometry Working Meeting, 14-18 March 2021 Cyberspace, Gothenburg, Sweden, Ed. R. Haas, ISBN: 978-91-88041-41-8, pp. 19-23
   \bibitem[\protect\citeauthoryear{Niell et al.}{2018}]{Niell2018} Niell A., Barrett J., Burns A., Cappallo R., Corey B., Derome M., Eckert C., et al., 2018, RaSc, 53, 1269. doi:10.1029/2018RS006617
   \bibitem[\protect\citeauthoryear{Nothnagel et al.}{2017}]{Nothnagel2017} Nothnagel A., Artz T., Behrend D., Malkin Z., 2017, JGeod, 91, 711. doi:10.1007/s00190-016-0950-5
   \bibitem[\protect\citeauthoryear{P{\'e}rez-D{\'\i}ez et al.}{2024}]{Perez-Diez2024} P{\'e}rez-D{\'\i}ez V., Mart{\'\i}-Vidal I., Albentosa-Ruiz E., Gonz{\'a}lez-Garc{\'\i}a J., Jaron F., Savolainen T., Xu M.~H., et al., 2024, A\&A, 688, A151. doi:10.1051/0004-6361/202348633
   \bibitem[\protect\citeauthoryear{Petrachenko et al.}{2009}]{Petrachenko2009} Petrachenko B., Niell A., Behrend D., Corey B., Boehm J., Charlot P., Collioud A., et al., 2009, NASA/TM-2009-214180
   \bibitem[\protect\citeauthoryear{Schartner et al.}{2023}]{Schartner2023} Schartner M., Collioud A., Charlot P., Xu M.~H., Soja B., 2023, JGeod, 97, 17. doi:10.1007/s00190-023-01706-4
   \bibitem[\protect\citeauthoryear{Sovers, Fanselow, \& Jacobs}{1998}]{Sovers1998} Sovers O.~J., Fanselow J.~L., Jacobs C.~S., 1998, RvMP, 70, 1393. doi:10.1103/RevModPhys.70.1393
\end{thebibliography}
\end{document}